\documentclass[aps,prl,twocolumn,showpacs,amsmath,amssymb]{revtex4-1}
\usepackage{amsmath} \usepackage{graphicx} \usepackage{subfigure}
\usepackage{epstopdf} \usepackage{color} \usepackage{multirow}
\usepackage{setspace} \usepackage{overpic} \usepackage{amssymb}
\usepackage{ulem}
\usepackage[bookmarksnumbered, pdfstartview=FitH,colorlinks,urlcolor=blue, citecolor=blue,linkcolor=blue] {hyperref}
\usepackage{lineno}
\usepackage{bm}
\usepackage{rotating}
\usepackage{xcolor}
\usepackage{makecell}
\usepackage{mathtext}
\usepackage{mathrsfs}
\usepackage{overpic}
\usepackage[T1]{fontenc}
\usepackage{lmodern}
\usepackage[utf8]{inputenc}

\let\oldequation\equation
\let\oldendequation\endequation
\renewenvironment{equation}
 {\linenomathNonumbers\oldequation}
 {\oldendequation\endlinenomath}
\newcommand{\nppi}{$\bar{n} p \to 2\pi^{+}\pi^{-}i\pi^{0}$}
\newcommand{\jnp}{$J/\psi \to p \pi^{-} \bar{n}$}

\newcommand{\BESIIIorcid}[1]{\href{https://orcid.org/#1}{\hspace*{0.1em}\raisebox{-0.45ex}{\includegraphics[width=1em]{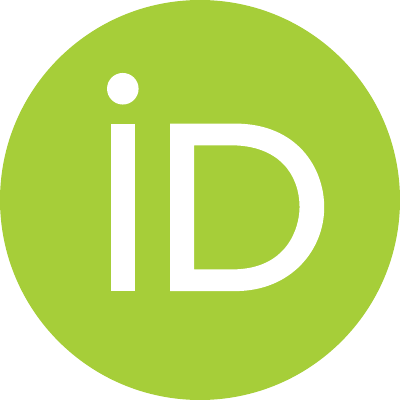}}}} 

\begin{document}
\definecolor{boslv}{rgb}{0.0, 0.65, 0.58}
\definecolor{Munsell}{HTML}{00A877}
\newcommand{\psip}{\psi^{'}}
\newcommand{\psipp}{\psi(3686)}

\newcommand{\Br}{\mathcal{B}}
\newcommand{\too}{\rightarrow}
\newcommand{\del}{\color{red}\sout}
\newcommand{\new}{\color{blue}\uwave}

\def\ifb{\mbox{fb$^{-1}$}}
\def\ipb{\mbox{pb$^{-1}$}}
\def\inb{\mbox{nb$^{-1}$}}
\newcommand{\lamcplamcm}{\Lambda_{c}^{+}\bar{\Lambda}_{c}^{-}}
\newcommand{\lambdacp}{\Lambda_{c}^{+}}
\newcommand{\lambdacm}{\bar{\Lambda}_{c}^{-}}
\newcommand{\afs}{\alpha_s}
\newcommand{\bgp}{\beta\gamma}
\newcommand{\eff}{\varepsilon}
\newcommand{\sintht}{\sin{\theta}}
\newcommand{\costht}{\cos{\theta}}
\newcommand{\dedx}{dE/dx}

\newcommand{\probfc}{Prob_{\chi^2}}
\newcommand{\probpi}{Prob_{\pi}}
\newcommand{\probka}{Prob_{K}}
\newcommand{\probpr}{Prob_{p}}
\newcommand{\proball}{Prob_{all}}

\newcommand{\chicJ}{\chi_{cJ}}
\newcommand{\gchicJ}{\gamma\chi_{cJ}}
\newcommand{\gchica}{\gamma\chi_{c0}}
\newcommand{\gchicb}{\gamma\chi_{c1}}
\newcommand{\gchicc}{\gamma\chi_{c2}}
\newcommand{\hc}{h_c(^1p_1)}
\newcommand{\qqb}{q\bar{q}}
\newcommand{\uub}{u\bar{u}}
\newcommand{\ddb}{d\bar{d}}
\newcommand{\SSB}{\Sigma^+\bar{\Sigma}^-}
\newcommand{\LLB}{\Lambda\bar{\Lambda}}
\newcommand{\ccb}{c\bar{c}}

\newcommand{\psipto}{\psi^{\prime}\rightarrow \pi^+\pi^- J/\psi}
\newcommand{\ptomm}{J/\psi\rightarrow \mu^+\mu^-}
\newcommand{\ppp}{\pi^+\pi^- \pi^0}
\newcommand{\pip}{\pi^+}
\newcommand{\pim}{\pi^-}
\newcommand{\kap}{K^+}
\newcommand{\kam}{K^-}
\newcommand{\ks}{K^0_s}
\newcommand{\pbar}{\bar{p}}
\newcommand{\jp}{J/\psi\rightarrow \gamma\pi^0}
\newcommand{\je}{J/\psi\rightarrow \gamma\eta}
\newcommand{\jep}{J/\psi\rightarrow \gamma\eta^{\prime}}

\newcommand{\LL}{\ell^+\ell^-}
\newcommand{\ee}{e^+e^-}
\newcommand{\MM}{\mu^+\mu^-}
\newcommand{\GG}{\gamma\gamma}
\newcommand{\TT}{\tau^+\tau^-}
\newcommand{\pp}{\pi^+\pi^-}
\newcommand{\kk}{K^+K^-}
\newcommand{\ppb}{p\bar{p}}
\newcommand{\gpp}{\gamma \pi^+\pi^-}
\newcommand{\gkk}{\gamma K^+K^-}
\newcommand{\gppb}{\gamma p\bar{p}}
\newcommand{\ggee}{\gamma\gamma e^+e^-}
\newcommand{\gguu}{\gamma\gamma\mu^+\mu^-}
\newcommand{\ggll}{\gamma\gamma l^+l^-}
\newcommand{\ppee}{\pi^+\pi^- e^+e^-}
\newcommand{\ppuu}{\pi^+\pi^-\mu^+\mu^-}
\newcommand{\etap}{\eta^{\prime}}
\newcommand{\gpi}{\gamma\pi^0}
\newcommand{\geta}{\gamma\eta}
\newcommand{\getap}{\gamma\etap}
\newcommand{\pppp}{\pi^+\pi^-\pi^+\pi^-}
\newcommand{\ppkk}{\pi^+\pi^-K^+K^-}
\newcommand{\pppr}{\pi^+\pi^-p\bar{p}}
\newcommand{\kkkk}{K^+K^-K^+K^-}
\newcommand{\kskp}{K^0_s K^+ \pi^- + c.c.}
\newcommand{\ppkp}{\pi^+\pi^-K^+ \pi^- + c.c.}
\newcommand{\ksks}{K^0_s K^0_s}
\newcommand{\dphi}{\phi\phi}
\newcommand{\phikk}{\phi K^+K^-}
\newcommand{\ppeta}{\pi^+\pi^-\eta}
\newcommand{\gpppp}{\gamma \pi^+\pi^-\pi^+\pi^-}
\newcommand{\gppkk}{\gamma \pi^+\pi^-K^+K^-}
\newcommand{\gpppr}{\gamma \pi^+\pi^-p\bar{p}}
\newcommand{\gkkkk}{\gamma K^+K^-K^+K^-}
\newcommand{\gkskp}{\gamma K^0_s K^+ \pi^- + c.c.}
\newcommand{\gppkp}{\gamma \pi^+\pi^-K^+ \pi^- + c.c.}
\newcommand{\gksks}{\gamma K^0_s K^0_s}
\newcommand{\gphiphi}{\gamma \phi\phi}

\newcommand{\tpp}{3(\pi^+\pi^-)}
\newcommand{\tppkk}{2(\pi^+\pi^-)(K^+K^-)}
\newcommand{\pptkk}{(\pi^+\pi^-)2(K^+K^-)}
\newcommand{\tkk}{3(K^+K^-)}
\newcommand{\gtpp}{\gamma 3(\pi^+\pi^-)}
\newcommand{\gtppkk}{\gamma 2(\pi^+\pi^-)(K^+K^-)}
\newcommand{\gpptkk}{\gamma (\pi^+\pi^-)2(K^+K^-)}
\newcommand{\gtkk}{\gamma 3(K^+K^-)}

\newcommand{\psp}{\psi(3686)}
\newcommand{\jpsi}{J/\psi}
\newcommand{\ar}{\rightarrow}
\newcommand{\lra}{\longrightarrow}
\newcommand{\jpsito}{J/\psi \rightarrow }
\newcommand{\ptoppjp}{J/\psi \rightarrow\pi^+\pi^- J/\psi}
\newcommand{\pspto}{\psi^\prime \rightarrow }
\newcommand{\ptop}{\psi'\rightarrow\pi^0 J/\psi}
\newcommand{\ptoeta}{\psi'\rightarrow\eta J/\psi}
\newcommand{\ecto}{\eta_c \rightarrow }
\newcommand{\ecpto}{\eta_c^\prime \rightarrow }
\newcommand{\xto}{X(3594) \rightarrow }
\newcommand{\chicJto}{\chi_{cJ} \rightarrow }
\newcommand{\chiczto}{\chi_{c0} \rightarrow }
\newcommand{\chicoto}{\chi_{c1} \rightarrow }
\newcommand{\chictto}{\chi_{c2} \rightarrow }
\newcommand{\pspp}{\psi^{\prime\prime}}
\newcommand{\ptochic}{\psi(2S)\ar \gamma\chi_{c1,2}}
\newcommand{\ppjpsi}{\pi^0\pi^0 J/\psi}
\newcommand{\utoeta}{\Upsilon^{\prime}\ar\eta\Upsilon}
\newcommand{\ww}{\omega\omega}
\newcommand{\wf}{\omega\phi}
\newcommand{\ff}{\phi\phi}
\newcommand{\npsp}{N_{\psp}}
\newcommand{\llb}{\Lambda\bar{\Lambda}}
\newcommand{\llbpi}{\llb\pi^0}
\newcommand{\llbeta}{\llb\eta}
\newcommand{\ppi}{p\pi^-}
\newcommand{\pbpi}{\bar{p}\pi^+}
\newcommand{\lamb}{\bar{\Lambda}}
\def\ctup#1{$^{\cite{#1}}$}
\newcommand{\bfg}{\begin{figure}}
\newcommand{\efg}{\end{figure}}
\newcommand{\bitm}{\begin{itemize}}
\newcommand{\eitm}{\end{itemize}}
\newcommand{\bnum}{\begin{enumerate}}
\newcommand{\enum}{\end{enumerate}}
\newcommand{\btbl}{\begin{table}}
\newcommand{\etbl}{\end{table}}
\newcommand{\btbu}{\begin{tabular}}
\newcommand{\etbu}{\end{tabular}}
\newcommand{\bcl}{\begin{center}}
\newcommand{\ecl}{\end{center}}
\newcommand{\bbt}{\bibitem}
\newcommand{\beq}{\begin{equation}}
\newcommand{\eeq}{\end{equation}}
\newcommand{\beqr}{\begin{eqnarray}}
\newcommand{\eeqr}{\end{eqnarray}}
\newcommand{\red}{\color{red}}
\newcommand{\blue}{\color{blue}}
\newcommand{\yellow}{\color{yellow}}
\newcommand{\green}{\color{green}}
\newcommand{\purple}{\color{purple}}
\newcommand{\brown}{\color{brown}}
\newcommand{\black}{\color{black}}

\title{\boldmath Study of the reactions $\bar{n} p \to 2\pi^{+}\pi^{-}$, $2\pi^{+}\pi^{-}\pi^{0}$, and $2\pi^{+}\pi^{-}2\pi^{0}$  using $J/\psi \to p \pi^{-}\bar{n}$ }
\author{
\begin{small}
\begin{center}
M.~Ablikim$^{1}$\BESIIIorcid{0000-0002-3935-619X},
M.~N.~Achasov$^{4,c}$\BESIIIorcid{0000-0002-9400-8622},
P.~Adlarson$^{81}$\BESIIIorcid{0000-0001-6280-3851},
X.~C.~Ai$^{86}$\BESIIIorcid{0000-0003-3856-2415},
R.~Aliberti$^{39}$\BESIIIorcid{0000-0003-3500-4012},
A.~Amoroso$^{80A,80C}$\BESIIIorcid{0000-0002-3095-8610},
Q.~An$^{77,64,\dagger}$,
Y.~Bai$^{62}$\BESIIIorcid{0000-0001-6593-5665},
O.~Bakina$^{40}$\BESIIIorcid{0009-0005-0719-7461},
Y.~Ban$^{50,h}$\BESIIIorcid{0000-0002-1912-0374}
% ...
\\
\vspace{0.2cm}
(BESIII Collaboration)\\
\vspace{0.2cm} {\it
$^{1}$ Institute of High Energy Physics, Beijing 100049, People's Republic of China\\
% ...
$^{4}$ Budker Institute of Nuclear Physics SB RAS (BINP), Novosibirsk 630090, Russia\\
% ...
$^{39}$ Johannes Gutenberg University of Mainz, Johann-Joachim-Becher-Weg 45, D-55099 Mainz, Germany\\
$^{40}$ Joint Institute for Nuclear Research, 141980 Dubna, Moscow region, Russia\\
% ...
$^{50}$ Peking University, Beijing 100871, People's Republic of China\\
% ...
$^{62}$ Southeast University, Nanjing 211100, People's Republic of China\\
% ...
$^{64}$ State Key Laboratory of Particle Detection and Electronics, Beijing 100049, Hefei 230026, People's Republic of China\\
% ...
$^{77}$ University of Science and Technology of China, Hefei 230026, People's Republic of China\\
% ...
$^{80}$ University of Turin and INFN, (A)University of Turin, I-10125, Turin, Italy; (B)University of Eastern Piedmont, I-15121, Alessandria, Italy; (C)INFN, I-10125, Turin, Italy\\
$^{81}$ Uppsala University, Box 516, SE-75120 Uppsala, Sweden\\
% ...
$^{86}$ Zhengzhou University, Zhengzhou 450001, People's Republic of China\\
\vspace{0.2cm}
$^{\dagger}$ Deceased\\
% ...
$^{c}$ Also at the Novosibirsk State University, Novosibirsk, 630090, Russia\\
% ...
$^{h}$ Also at State Key Laboratory of Nuclear Physics and Technology, Peking University, Beijing 100871, People's Republic of China\\
% ...
}\end{center}
\vspace{0.4cm}
\end{small}
}

\date{\today}


\begin{abstract}
We report an experimental investigation of the reactions $\bar{n} p \to 2\pi^{+}\pi^{-}$, $\bar{n} p \to 2\pi^{+}\pi^{-}\pi^{0}$, and $\bar{n} p \to 2\pi^{+}\pi^{-}2\pi^{0}$  using $(10.087 \pm 0.044) \times 10^{9}$ $J/\psi$ events collected with the BESIII detector at the BEPCII storage ring. The antineutron ($\bar{n}$) is produced in the decay $J/\psi \to p \pi^{-} \bar{n}$ with  studied momentum from 200~MeV/$c$ to 1174~MeV/$c$, while the target proton originates from the hydrogen nuclei in the cooling oil of the beam pipe.
This novel method pioneers the study of $\bar{n}$-nucleon interactions at an $e^{+}e^{-}$ collider, providing the first experimental data for $\bar{n}$ momenta exceeding 800~MeV/$c$.

\end{abstract}

\maketitle

Scattering experiments are fundamental tools for probing the internal structure of nucleons, with the  choice of particle source playing a pivotal role. Antineutrons possess a sufficiently long lifetime to serve as a viable particle source. However, since their discovery~\cite{Cork:1957nu}, antineutrons have rarely been utilized due to the significant challenges associated with their production and control. Despite these difficulties, antineutron beams offer unique advantages, such as the absence of Coulomb corrections and the purity of the $I = 1$ state in the $\bar{n}p$ system, which simplifies the analysis of isospin-dependent effects~\cite{allReview}.

To date, the most effective antineutron sources  have been achieved at the BNL E-767 experiment~\cite{BNL}, with momenta ranging from 100~MeV/$c$ to 500~MeV/$c$, and the CERN OBELIX experiment~\cite{OBELIX}, with momenta ranging from 50~MeV/$c$ to 400~MeV/$c$.  
Both facilities produced antineutron sources through the charge-exchange  process $\bar{p}p \to n \bar{n}$. Despite the experimental insights obtained, this method has notable drawbacks: the antineutron production rates remain low, and achieving precise control over their momenta and directions poses significant technical challenges. 
Significantly, experimental data for $\bar{n}p$ annihilation cross sections are currently non-existent for antineutron momenta exceeding 800~MeV/$c$. This higher-momentum region, extending up to 1174~MeV/$c$ in this study, represents a critical transition zone where the relevant degrees of freedom shift from mesons and nucleons to quarks and gluons in Quantum Chromodynamics. Our work provides the first measurements in this previously unexplored energy regime, filling a long-standing experimental vacuum and marking the first such observation at an $e^+e^-$ collider. These results offer unique and essential benchmarks for the development of future $N\bar{N}$ interaction models.

In this Letter, we use a novel source of antineutrons~\cite{ycz} from $J/\psi$ decays to overcome traditional production challenges. The $J/\psi$ meson, produced copiously in $\ee$ annihilation with a branching fraction of \jnp, $(2.12\pm0.09)\times10^{-3}$~\cite{pdg2024}, emerges as a promising new source of $\bar{n}$. 
By measuring $p\pi^{-}$, we can accurately determine the momentum of the $\bar{n}$.
While previous BESIII studies~\cite{BES_scatter1, BES_scatter2, BES_scatter3} used two-body $J/\psi$ decays for fixed-momentum hyperon ($\Lambda, \Sigma, \rm{and}~ \Xi$) scattering, the three-body \jnp~decay here provides a continuous antineutron spectrum, enabling a cross-section scan over a wide momentum range.
The interaction between the $\bar{n}$ and   protons in the target material, located within a  small  radius of the $\ee$ beam pipe, enables a  precise study of the  cross section and the properties of the final states.
Following the single-tag (ST) and double-tag (DT) approach described in Refs.~\cite{BES_scatter3, BES_scatter33}, we employ a similar analysis strategy to minimize systematic uncertainties associated with the  branching fraction of \jnp, the $p\pi^{-}$ selection, and the total number of $J/\psi$ events. 
Specifically, the ST selection involves reconstructing $p\pi^{-}$ from \jnp. The DT selection extends this by additionally reconstructing the final-state particles from the   $\bar{n}p$ interaction, $2\pi^{+}\pi^{-}i\pi^{0}$ ($i=0, 1, 2$), as  discussed in detail later.


A total of $(10.087 \pm 0.044) \times 10^{9}$ $J/\psi$ events collected by the BESIII detector at the center-of-mass energy 3.097~GeV are used in this analysis.  The BESIII detector records symmetric $e^{+}e^{-}$ collisions at the BEPCII collider~\cite{CXYu_bes3}. Detailed information about the BESIII detector can be found in Ref.~\cite{Ablikim:2009aa}. 
The BESIII detector surrounds the beam pipe, which consists of four layers: a gold ($^{197}$Au) layer, a beryllium ($^{9}$Be) layer, an oil layer ($^{12}$C, $^{1}$H) serving as a coolant, and another beryllium layer. The distances from the inner and outer surfaces of the oil layer to the beam direction are 3.23~cm and 3.31~cm, respectively.
Since the proton in the $^{1}$H of the cooling oil is essentially at rest, whereas the protons in the $^{197}$Au/$^{9}$Be/$^{12}$C nuclei have Fermi motion, our focus is solely on the protons from the $^{1}$H. By utilizing these almost static protons in the cooling oil of the beam pipe, we can directly obtain information on the antineutron-proton interaction. This approach is similar to that adopted in Ref.~\cite{BES_scatter2}, which studied the $\Lambda p$ reaction by utilizing the proton from the cooling oil.


 Simulated samples produced with a {\sc geant4}-based~\cite{geant4} Monte Carlo (MC) package, which includes the geometric description~\cite{detvis} of the BESIII detector and the detector response, are used to optimize the event selection criteria and estimate the signal efficiency and backgrounds. The simulation models the beam-energy spread and initial-state radiation in the $e^+e^-$ annihilation using the generator {\sc kkmc}~\cite{kkmc_b}. 
The inclusive MC sample  includes the production of the $J/\psi$ resonance and the continuum processes incorporated in {\sc kkmc}. All particle decays are modeled with {\sc evtgen}~\cite{evtg1,evtg2} using branching fractions either taken from the Particle Data Group (PDG)~\cite{pdg2024},  where available, or otherwise estimated  with    {\sc lundcharm}~\cite{lundcharm_a,lundcharm_b}.   
Final state radiation  from charged  particles is incorporated using the  {\sc photos} package~\cite{photos}.
To estimate the detection efficiency, we generate  one million  \jnp~ST MC events,  and about 2.3 million \jnp~with \nppi~events for each signal process as the DT MC samples.
The  \nppi~processes are simulated by assuming the proton to be at rest~\cite{BES_scatter2} and modeled with a uniform phase space distribution.
In the signal MC simulation, the momentum and angular distributions of $\bar{n}$ are generated according to a  partial-wave analysis (PWA)  conducted on \jnp.
The PWA utilizes a high-purity selected data sample of $10^{5}$  events, and uses the open-source TF-PWA framework~\cite{TF_pwa}, employing  the helicity amplitude formalism~\cite{PWA_model1,PWA_model2}.  All  known $N^{\ast}$ resonances from PDG are incorporated for $p\pi^{-}$ and $\bar{n}\pi^{-}$, each modeled by relativistic Breit-Wigner functions, with masses and widths  fixed to the world averages~\cite{pdg2024}.
Figure~\ref{fig1} shows the  distributions of the momentum and cos$\theta$ of the $\bar{n}$, 
with a $z$-axis corresponding to the symmetry axis of the multilayer drift chamber (MDC).
The good data-MC consistency  validates the reliability of the PWA result.

The cross sections for each signal process are measured in five different  $\bar{n}$ momentum intervals,  defined from 200~MeV/$c$ to 1174~MeV/$c$ with a step size of 200~MeV/$c$.

 \begin{figure}[htbp]
\begin{center}
\begin{minipage}[t]{1.0\linewidth}
\includegraphics[width=0.49\textwidth]{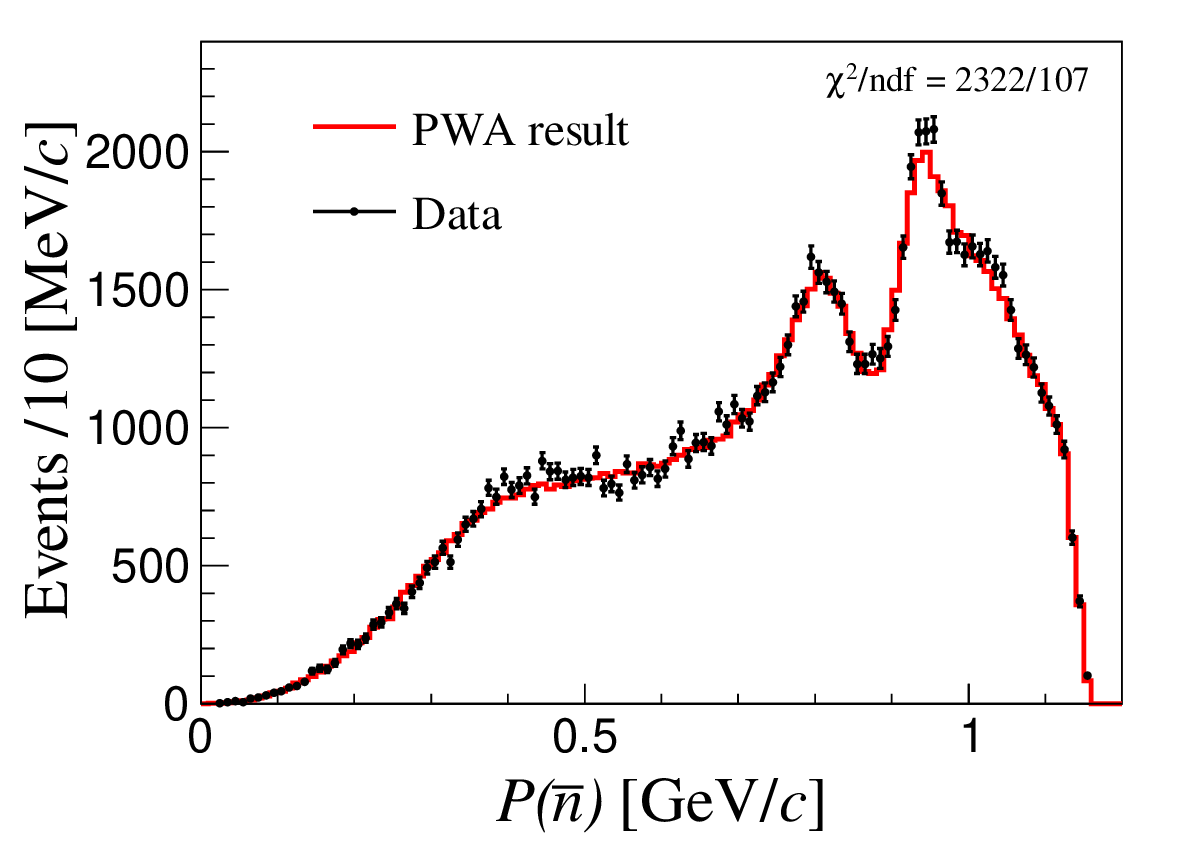}
\includegraphics[width=0.49\textwidth]{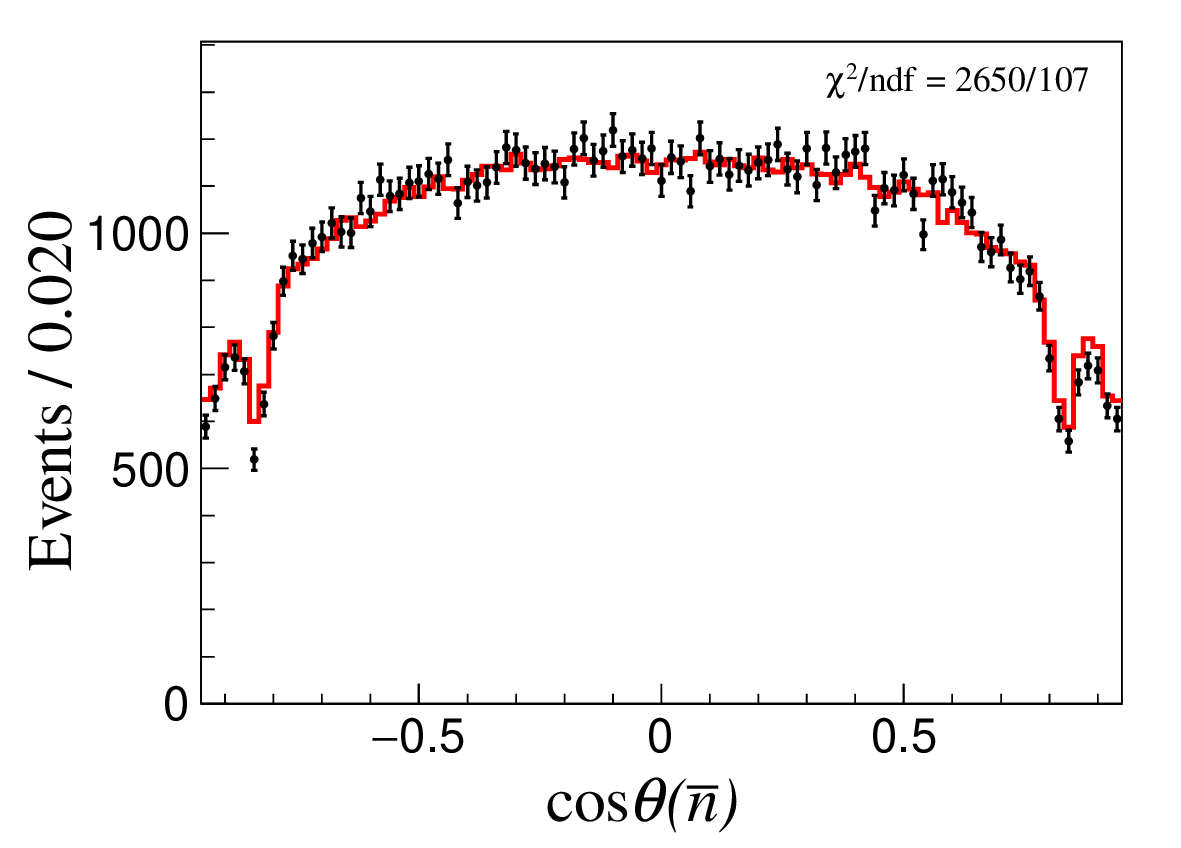}
\end{minipage}
\caption{The $P(\bar{n})$ (left) and cos$\theta(\bar{n})$ (right)  distributions for \jnp. The black dots with error bars  indicate data. The red histogram shows the PWA result. } \label{fig1}
\end{center}
\end{figure}


In the ST selection, candidate events  for \jnp ~are reconstructed by identifying $p\pi^{-}$ combinations. 
Charged tracks reconstructed in the main  drift chamber (MDC) are required to fall within the polar angle range $|\cos\theta'| < 0.93$, where $\theta'$ is the angle relative to the $z$-axis. 
Additionally, the distance of closest approach to the $\ee$ interaction point (IP) must be less than $\pm10$~cm along the $z$-axis and less than 1~cm in the transverse plane.
Particle identification (PID) for charged tracks is based on the combination of energy loss measurements $(dE/dx)$ in the MDC and flight time from the time-of-flight system. These measurements are used to construct likelihoods $\mathcal{L}(h)$ for each hadron $h$ hypothesis  ($h = p, K, \pi$), and  tracks are assigned the particle type corresponding to the highest likelihood.
At least two charged tracks, identified as a proton and a pion ($p\pi^{-}$), are required. No additional requirements are imposed on other charged tracks or neutral showers in the event. All valid $p\pi^{-}$ combinations are subjected to a vertex fit to improve track momentum resolution, followed by a four-constraint(4C) kinematic fit.
The 4C fit constrains the  missing track's mass  to  that of  the $\bar{n}$ and enforces energy-momentum conservation. 
For events with multiple candidates, the combination with the smallest sum of  $\chi^{2}_{\text{vertex}} + \chi^{2}_{\text{kinematic}}$ is  selected as the best candidate. This best combination is then required to satisfy $\chi^{2}_{\text{vertex}} < 10$ and $\chi^{2}_{\text{kinematic}} < 200$.
A relatively loose $\chi^{2}_{\text{kinematic}}$ requirement is chosen to preserve a linear background shape in the distribution of the recoil mass $RM(p\pi^{-}) = |p_{e^+e^-} - p_{p} - p_{\pi^{-}}|/c$, where $p_{e^+e^-}$, $p_{p}$, and $p_{\pi^{-}}$ are the four-momenta of the $e^+e^-$ initial state, proton, and pion, respectively. 
The  ST signal yield, $N_{\rm{ST}}$, is then extracted by fitting this $RM(p\pi^-)$ spectrum with a binned extended maximum likelihood fit. The $\bar{n}$ signal region is defined as $[0.92, 0.96]$~GeV/$c^2$.
The signal is modeled using a double-Gaussian function, and  the background is modeled with   a first-order Chebyshev  polynomial function. 
The fit result is shown in Fig.~\ref{fig2}. 
This fitting procedure is also applied independently within each $P(\bar n)$ momentum interval.
The ST selection efficiency, $\epsilon_{\rm{ST}}$, is determined from a dedicated ST Monte Carlo sample using the same reconstruction and selection criteria. The resulting values for $N_{\rm{ST}}$ and $\epsilon_{\rm{ST}}$ are summarized in Table~\ref{table-ST}.

\begin{figure}[htbp]
\begin{center}
\begin{minipage}[t]{1.0\linewidth}
\includegraphics[width=0.90\textwidth]{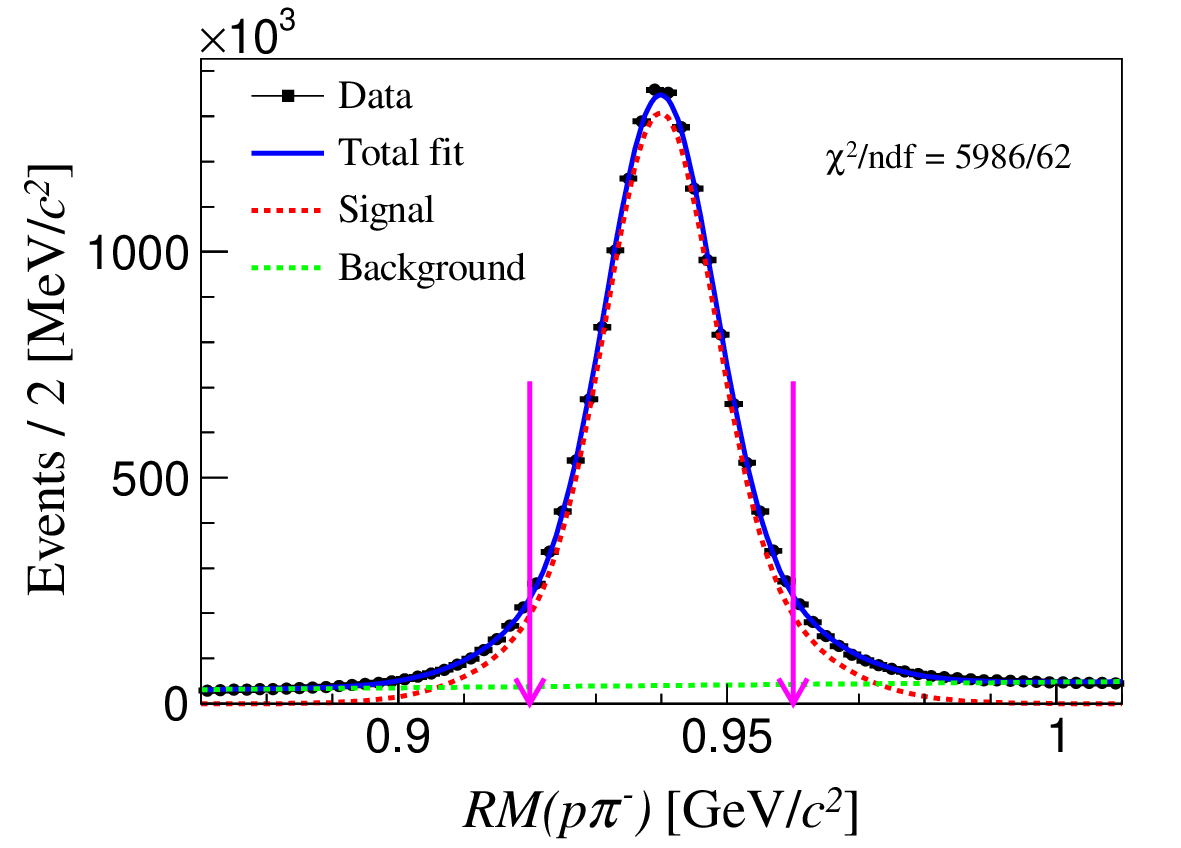}
\end{minipage}
\caption{The $RM(p\pi^{-})$  distribution with  $P(\bar{n})\in$ [200, 1174]~MeV/$c$.  The black dots with error bars represent data. The blue solid line is the total fit. The dashed red line is the ST signal, and the dashed green line is the background. The pink arrows indicate the selected signal range.} \label{fig2}
\end{center}
\end{figure}

\begin{table}[!htb]
\caption{The  sum of  $N_{\rm{ST}}$ and $\epsilon_{\rm{ST}}$ of the full sample and in different $P(\bar{n})$ intervals. }\label{table-ST}
\centering
\begin{tabular}{c|c|c}
\hline \hline
       $P(\bar{n})$~(MeV/$c$)       &  $N_{\rm{ST}}$~($\times 10^{3}$)      & $\epsilon_{\rm{ST}}$~(\%)     \\  \hline
 $[200, 1174]$        &  14863.8 $\pm$ 4.6  & 66.65                \\  \hline
 $[200, 400]$         &  1312.6 $\pm$ 1.3   & 75.30                 \\ 
 $[400, 600]$         &  2270.4 $\pm$ 1.7   & 73.35                 \\ 
 $[600, 800]$         &  3268.7 $\pm$ 2.1   & 69.28                 \\ 
 $[800, 1000]$        &  4863.1 $\pm$ 2.8   & 63.35                 \\ 
 $[1000, 1174]$       &  3166.6 $\pm$ 2.4   & 61.50                 \\ \hline \hline
\end{tabular}
\end{table}

In DT,  in addition to $p\pi^{-}$, the final state particles of  $2\pi^{+}\pi^{-}i\pi^{0}$ with $\pi^{0} \to \gamma \gamma$ are reconstructed.
The selected $p\pi^{-}$  combination is further required to satisfy $\chi^{2}_{\text{kinematic}} <5$. Because this results in a non-linear  background shape in the $RM(p\pi^{-})$ distribution, it is not used to determine the DT signal yield.
Additionally, we require three charged tracks, $2\pi^+\pi^-$, to satisfy the same criteria as the $\pi^{-}$ in ST. However, no requirement is imposed on the closest approach distance to IP, as $2\pi^{+}\pi^{-}$ originates from the beam pipe. 
A vertex fit is conducted for $2\pi^{+}\pi^{-}$, where the vertex position is denoted  as $(x_{1}, y_{1}, z_{1})$, and $\chi^{2}_{\text{vertex}}$ is required to be less than 5.
The direction of the $\bar{n}$ is determined from the $p\pi^{-}$ kinematic fit, with its intersection point on the beam pipe denoted as $(x_{2}, y_{2}, z_{2})$. 
The vector $(x_{1}-x_{2}, y_{1}-y_{2}, z_{1}-z_{2})$ is defined to utilize the relative positions of these two vertices, with signal events expected to be close to $(0, 0, 0)$.    
The requirements, $|\sqrt{(x_{1}^{2}+y_{1}^{2})} - \sqrt{(x_{2}^{2}+y_{2}^{2})}| < 0.5~\text{cm}$   and  $ |z_{1}-z_{2}| < 2.5~\text{cm}$, are applied.
For $\pi^0$ candidates, all  photon pairs are looped through with their invariant mass required to be within $(0.115, 0.150)$~GeV/$c^2$. A 1C kinematic fit, constraining the photon-pair mass to the known $\pi^0$ mass~\cite{pdg2024}, is then applied to improve the momentum resolution. For the three signal processes, at least $i$ $\pi^{0}$ candidate(s) must be identified. If there are additional $\pi^0$ candidates, those with the minimal  kinematic fit $\chi^2$ are selected. For the $i=2$ case, the two selected $\pi^0$ candidates  must not share any photons.

To remove  events from the reactions between  $\bar{n}$  and $^{197}$Au/$^{9}$Be/$^{12}$C nuclei,  we define the  momentum of the proton in the $^{1}$H of the cooling oil as $P(p_\text{oil}) = |\vec{P}(2\pi^+\pi^-i\pi^0)-[\vec{P}(\ee)-\vec{P}(p)-\vec{P}(\pi^{-})]|$, where $\vec{P}$  represents the three-momentum of each particle in the lab frame. 
As shown in Fig.~\ref{fig:p_oil}, $P(p_\text{oil})$ is negligible for signal processes but hundreds of  MeV/$c$ for the background processes where protons originate from atomic nuclei with Fermi motion.
To remove these background events, the requirements $P(p_\text{oil}) <$ 0.035~GeV/$c$, 0.040~GeV/$c$, and 0.045~GeV/$c$ are applied for $i = 0, 1$, and 2, respectively, optimized  using a figure of merit,  $S/\sqrt{S+B}$,  where $S$ in the numerator is the signal yield from the DT MC and $S+B$ in the denominator is the number of events from the data sample.

\begin{figure}[htbp]
    \begin{center}
    \begin{minipage}[t]{1.0\linewidth}
    \includegraphics[width=0.32\textwidth]{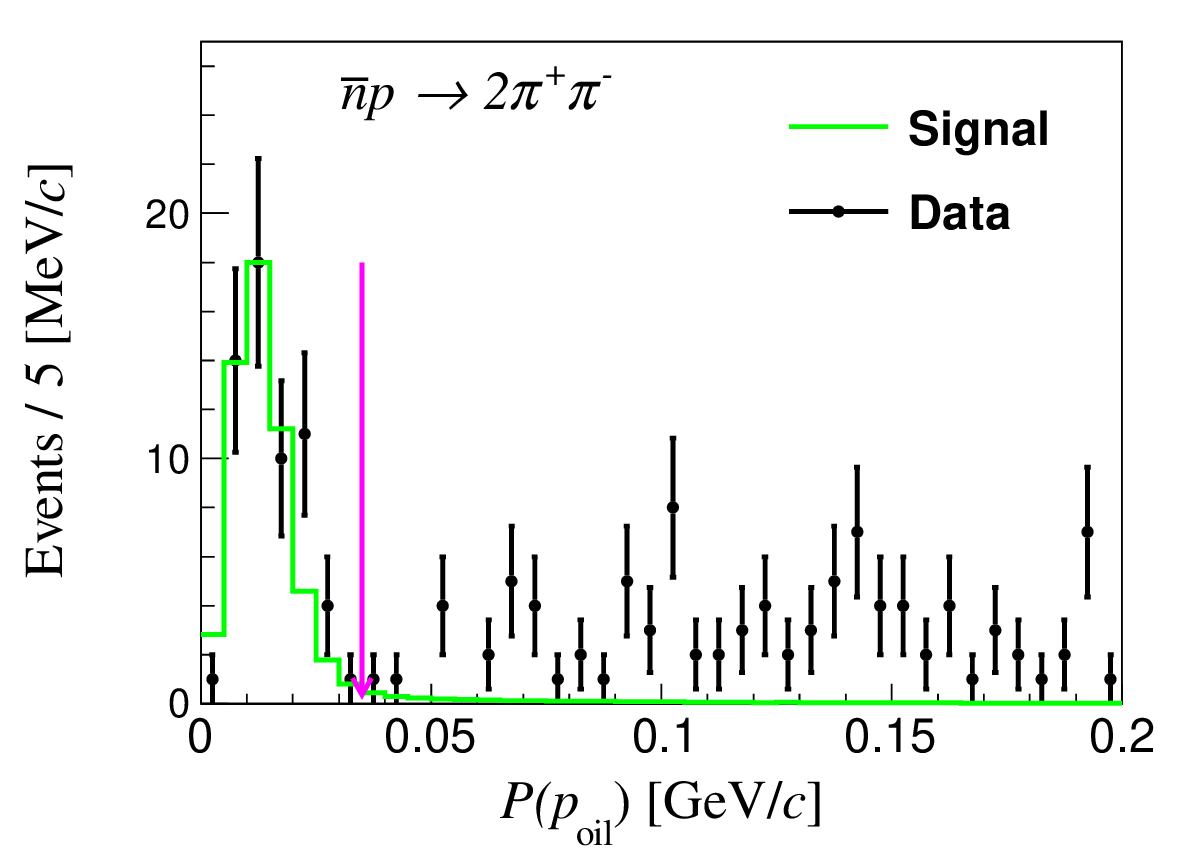}
    \includegraphics[width=0.32\textwidth]{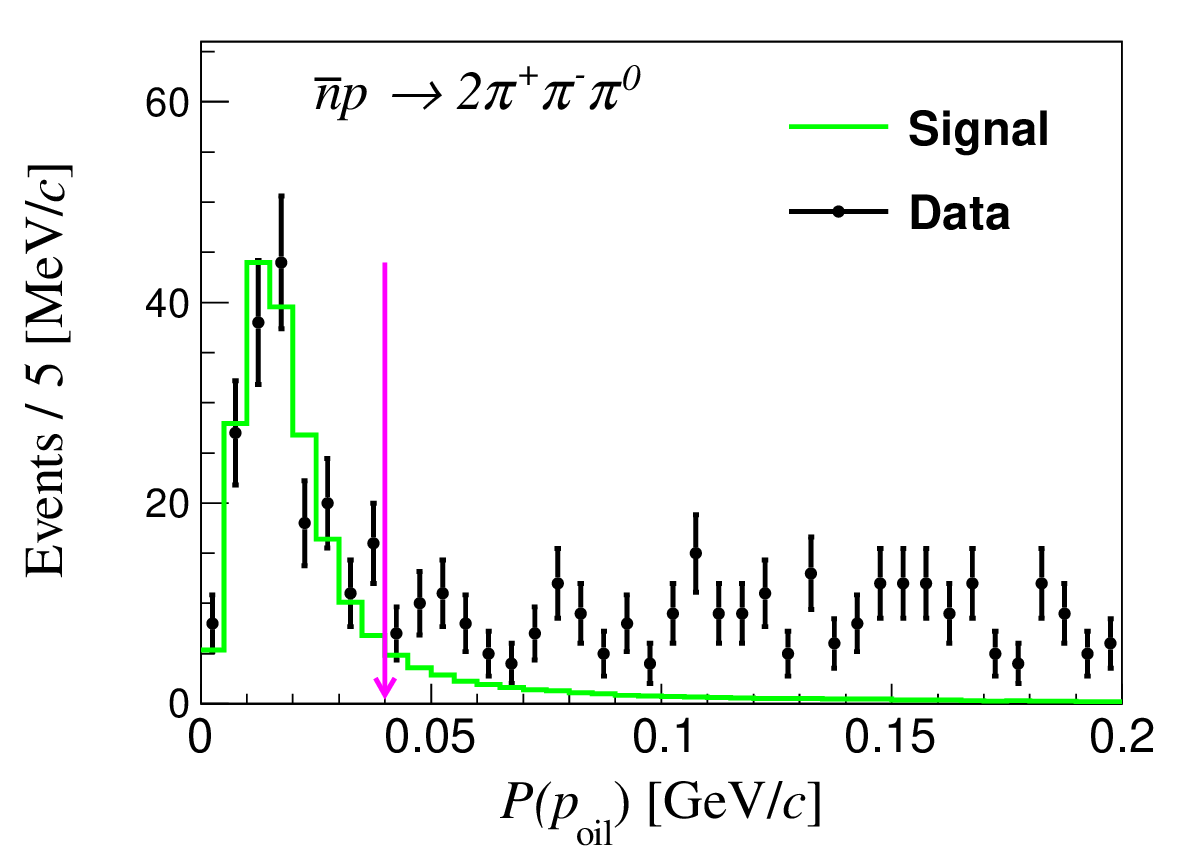}
    \includegraphics[width=0.32\textwidth]{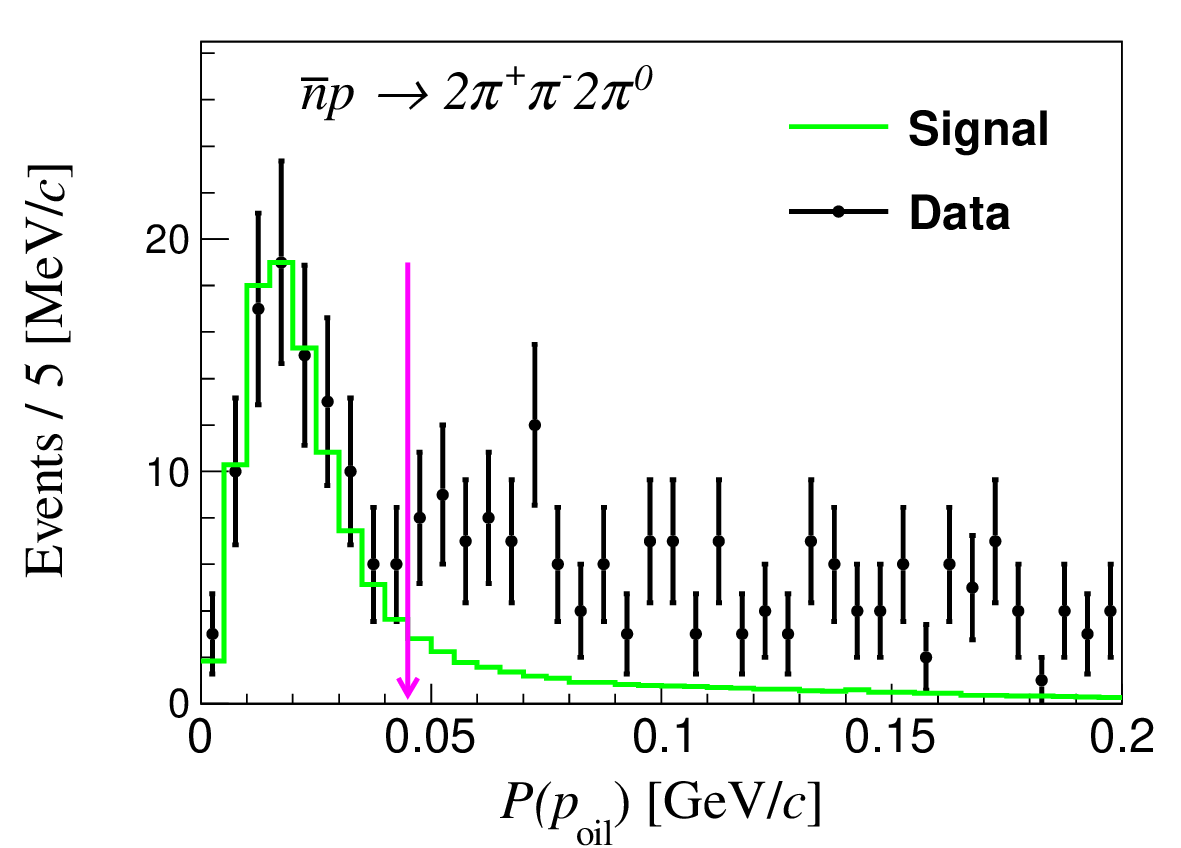}
    \end{minipage}
    \caption{Distributions of $P(p_\text{oil})$ for $\bar{n} p \to 2\pi^{+}\pi^{-}$(left), $\bar{n} p \to 2\pi^{+}\pi^{-}\pi^{0}$(middle), and $\bar{n} p \to 2\pi^{+}\pi^{-}2\pi^{0}$(right). The black dots with error bars represent the data, and the green lines are the signal MC. The pink arrows indicate the selected signal range.} \label{fig:p_oil}
    \end{center}
\end{figure}

For the signal process \nppi, the energy of  $\bar{n}$ and a static $p$   is  equivalent to that of $2\pi^{+}\pi^{-}i\pi^{0}$. Therefore, we define $\Delta E$ as the energy of $2\pi^{+}\pi^{-}i\pi^{0}$ after subtracting the energy of $\bar{n}$ and a static $p$. 
The number of DT events ($N_{\rm{DT}}$) is determined  by fitting the  $\Delta E$ distribution in data, 
 which includes both  $p\pi^{-}$ from the $J/\psi$ decay and $2\pi^{+}\pi^{-}i\pi^{0}$  from $\bar{n}p$ interaction, as given by:
\begin{equation}\label{eq1}
 N_{\rm{DT}} = \mathcal{L}(\bar{n}) \cdot \sigma_{i} \cdot \mathcal{B}^{i}_{\pi^0\to\gamma\gamma} \cdot \epsilon_{\rm{DT}},
\end{equation}
where $\sigma_{i}$ is the cross section of the processes \nppi, $\mathcal{B}^{i}_{\pi^0\to\gamma\gamma}$ is the branching fraction for $\pi^{0}\to\gamma\gamma$~\cite{pdg2024}  raised to the power $i$, and $\epsilon_{\rm{DT}}$ denotes the efficiency of the DT reconstruction.
The ``effective luminosity'' $\mathcal{L}(\bar{n})$ represents the integrated interaction opportunity between the incident $\bar{n}$ and the $^{1}$H target~\cite{lum_n}. It is determined by the number of produced $\bar{n}$ from $J/\psi \to p \pi^- \bar{n}$ decays and the target's physical properties, calculated as:
\begin{equation}\label{eq2}
  \mathcal{L}(\bar{n}) = \frac{N_{\rm{ST}}}{\epsilon_{\rm{ST}}}  \cdot \frac{N_{A} \cdot \rho \cdot l}{M},
\end{equation}
where $N_{A}$ is Avogadro's number~\cite{avod} and $l = 0.106$~cm is the average path length of $\bar{n}$ inside the oil layer, determined by averaging $t/\sin\theta(\bar{n})$ over the $\bar{n}$ angular distribution from PWA, with $t$ denoting the oil layer thickness. 
The cooling oil has a density of $\rho = 0.81$~g$\cdot$cm$^{-3}$ and a $^{12}$C : $^{1}$H mass fraction ratio of 84.9\% : 15.1\%~\cite{BES_scatter3}. Since the signal process specifically involves interactions with protons in the $^{1}$H component, an effective molar mass of $M = 6.634$~g$\cdot$mol$^{-1}$~\cite{mola} is adopted for the $^{1}$H target in Eq.~(\ref{eq2}).
Combining Eq.~(\ref{eq1}) and Eq.~(\ref{eq2}), the cross section $\sigma_{i}$ is determined as:
\begin{equation}\label{eq3}
   \sigma_{i} = \frac {M}{N_A\cdot\rho \cdot l}\cdot \frac{N_{\rm{DT}}} { \epsilon_{\rm{DT}} \cdot N_{\rm{ST}}/\epsilon_{\rm{ST}}  } \cdot \frac1{\mathcal{B}^{i}_{\pi^0\to\gamma\gamma}}. 
\end{equation}


Figure~\ref{fig3} shows the $\Delta E$ distributions from data after the final event selection. Clear enhancements  around 0~GeV/$c$, corresponding to the reactions \nppi, are observed.
Inclusive and signal MC studies confirm that peaking backgrounds and cross-feeds among different $\pi^0$ multiplicities are negligible due to kinematic constraints.
An unbinned  extended maximum likelihood fit is performed to the $\Delta E$ distribution to determine the signal yield.
We use the MC-determined shape convolved with a Gaussian function with free parameters to describe the signal shape. 
The background is described by a first-order Chebyshev  polynomial function. 
In the fit, the signal yield and the parameters of the Gaussian and  polynomial functions are free to vary. 
The fit results are shown in Fig.~\ref{fig3}.
The central values of  $\Delta E$ are negatively shifted mainly due to energy leakage during  $\pi^{0}$ reconstruction.
The same method was used to extract the  cross sections  in the  different $P(\bar{n})$   intervals.
Due to limited statistics, the Gaussian parameters are fixed to those obtained from the fit using the full $\bar{n}$  momentum range [200, 1174]~MeV/$c$ to ensure fit stability across all bins.
All the $N_{\rm{DT}}$ and $\epsilon_{\rm{DT}}$ values, and the PWA-weighted average momentum $\langle P \rangle$ to better characterize each momentum range, are  shown in Table~\ref{table3}.

\begin{figure*}[!htpb]
\begin{center}
\includegraphics[width=0.325\textwidth]{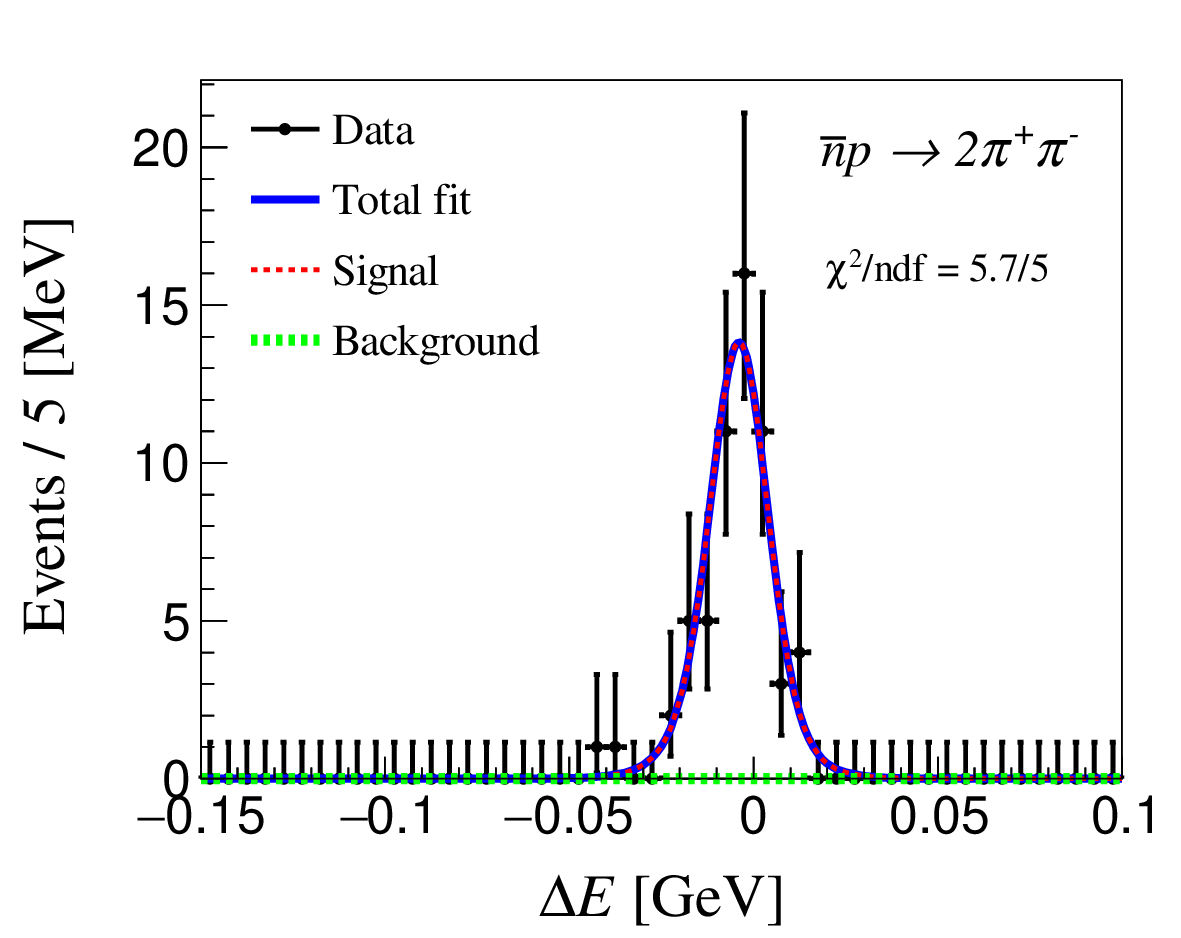}
\includegraphics[width=0.325\textwidth]{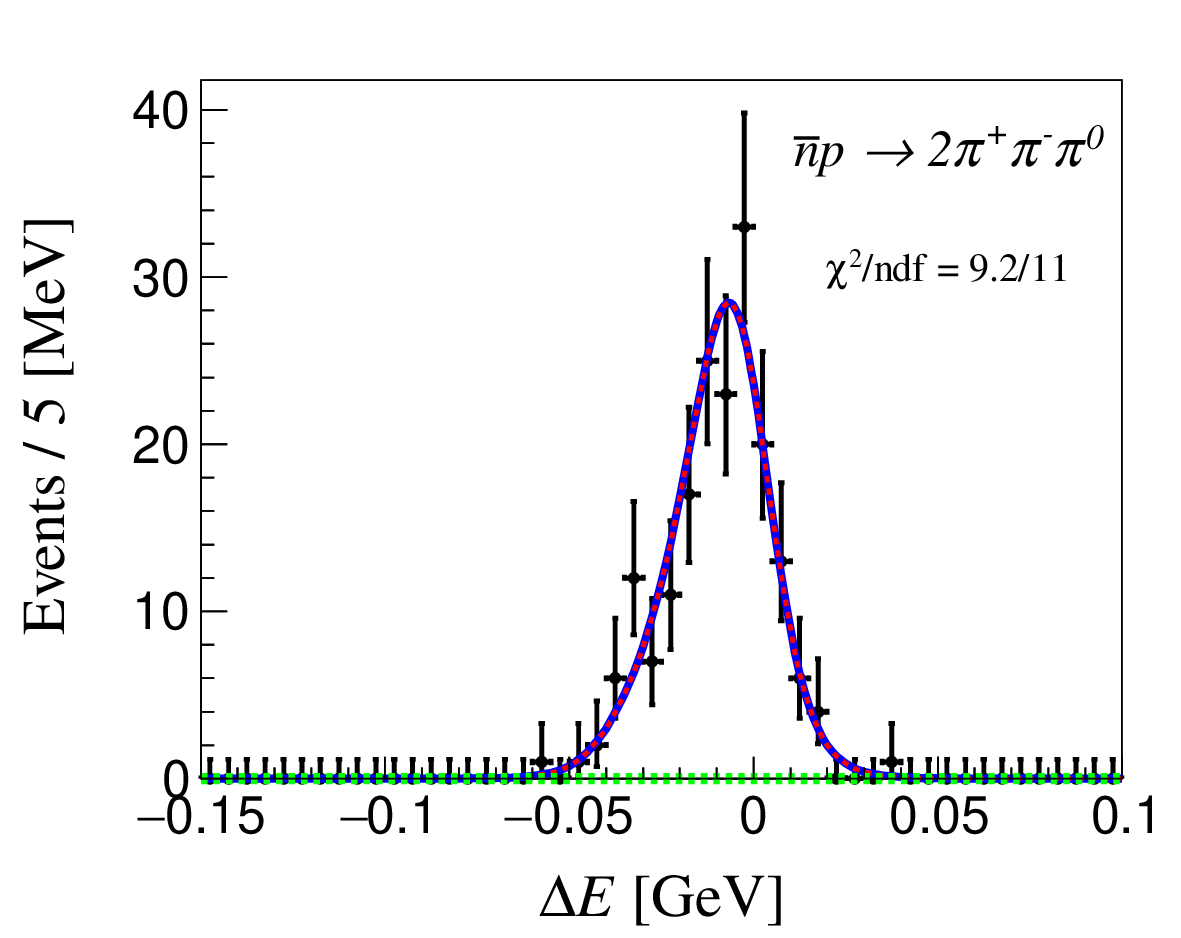}
\includegraphics[width=0.325\textwidth]{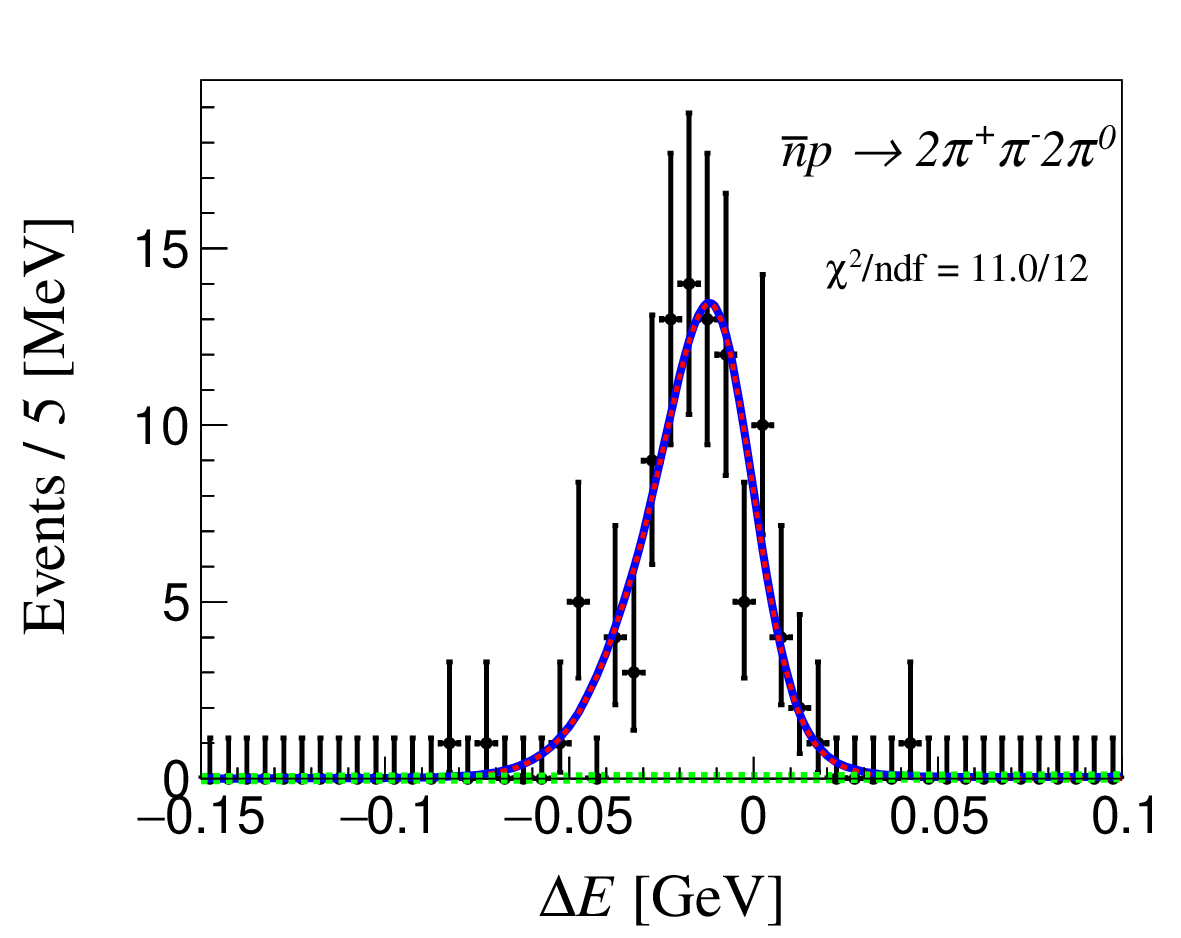}
\caption{The $\Delta E$  distributions with  $P(\bar{n})\in$~[200, 1174]~MeV/$c$.  The black dots with error bars represent the data. The blue solid lines are the total fits. The dashed red lines are the signals, and the dashed green lines are the backgrounds.} \label{fig3}
\end{center}
\end{figure*}

Furthermore, we examine the invariant mass distributions of the final state pions in two-body and three-body combinations of the events, as shown in Fig.~\ref{figmass}.
All possible combinations are included for each event. We  observe clear signals from $\rho$ and $\omega$ intermediate  states,  indicating significant contributions from these vector mesons.

Further studies with higher statistics and advanced techniques, including PWA to disentangle the $S$-wave and $P$-wave contributions,  would be crucial for quantifying the role of these intermediate states in mediating the antineutron-proton interaction.

\begin{figure}[htbp]
\begin{center}
\begin{minipage}[t]{1.0\linewidth}
\includegraphics[width=0.49\textwidth]{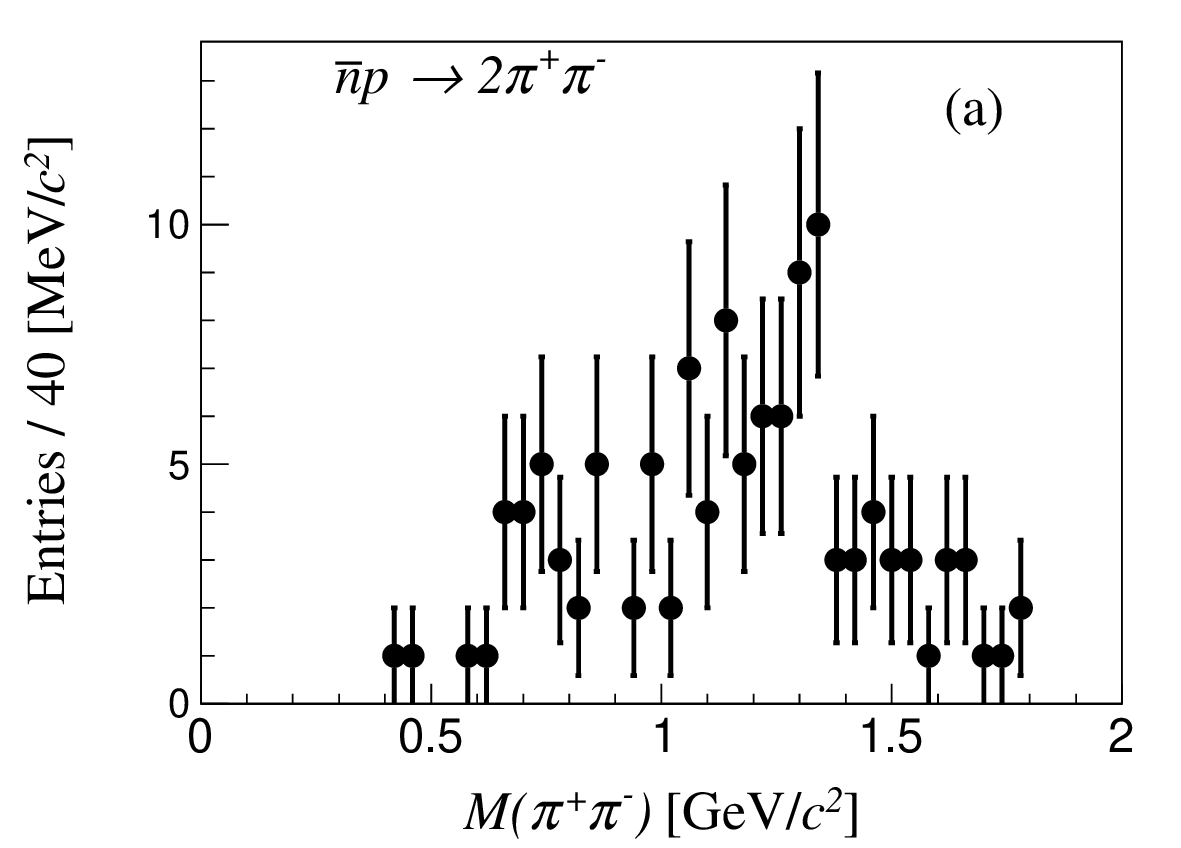}
\includegraphics[width=0.49\textwidth]{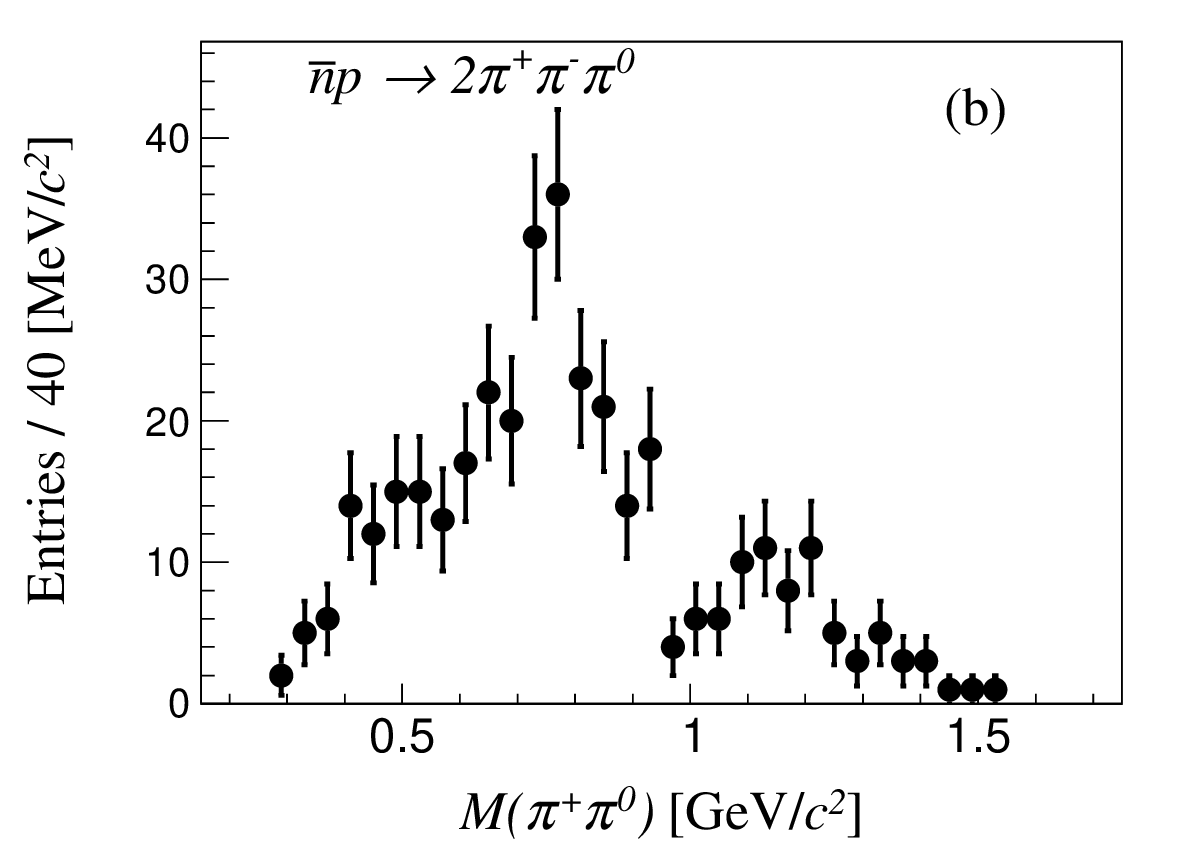}
\includegraphics[width=0.49\textwidth]{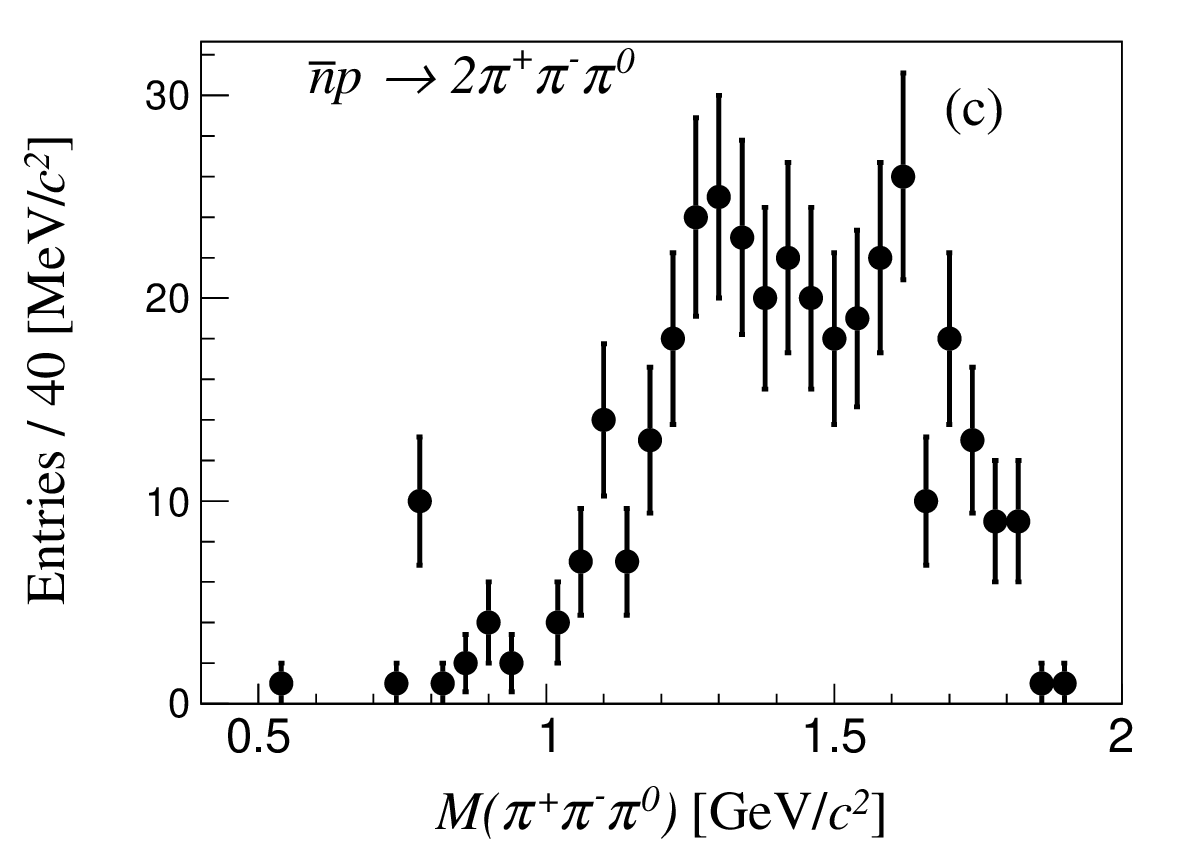}
\includegraphics[width=0.49\textwidth]{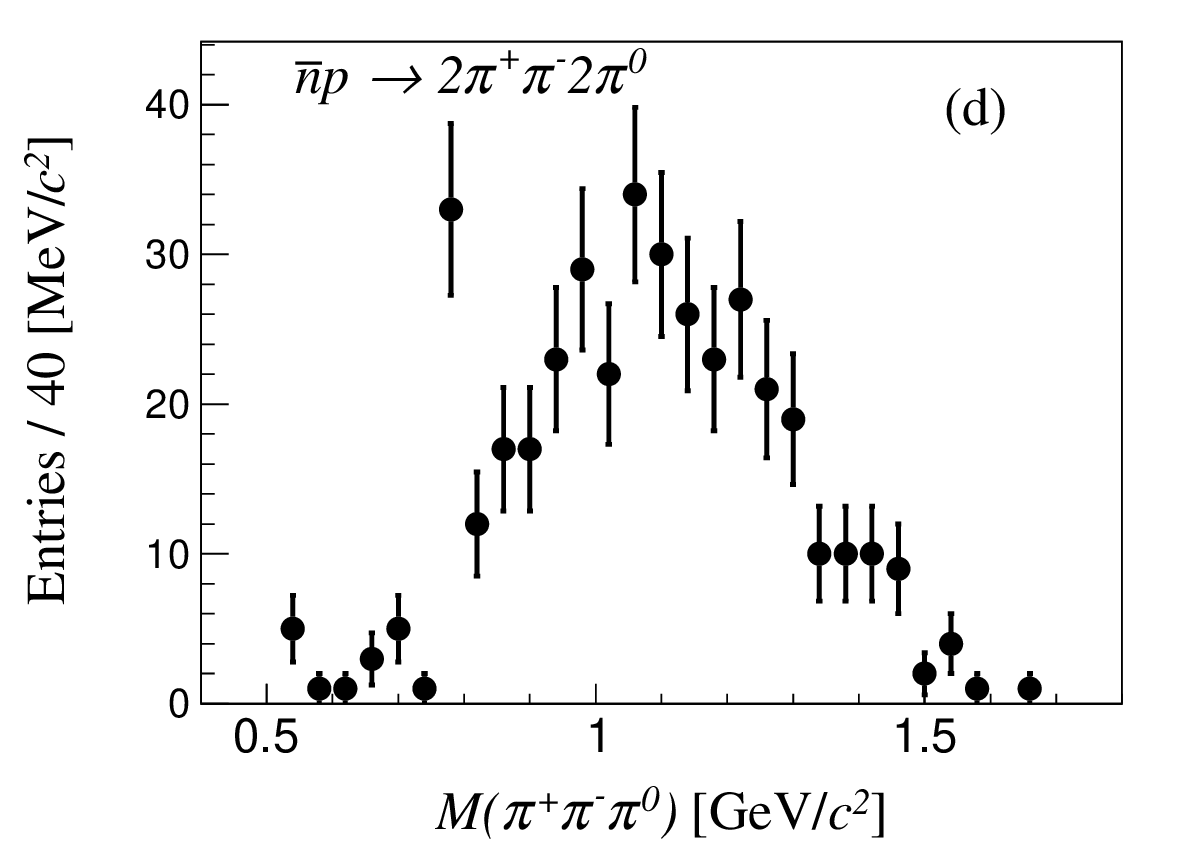}
\end{minipage}
\caption{Invariant mass distributions for (a) $M(\pi^{+}\pi^{-})$ in $\bar{n} p \to 2\pi^{+}\pi^{-}$; (b) $M(\pi^{+}\pi^{0})$ and (c) $M(\pi^{+}\pi^{-}\pi^{0})$ in $\bar{n} p \to 2\pi^{+}\pi^{-}\pi^{0}$; and (d) $M(\pi^{+}\pi^{-}\pi^{0})$ in $\bar{n} p \to 2\pi^{+}\pi^-2\pi^{0}$ for data.} \label{figmass}
\end{center}
\end{figure}

The cross section for each signal process is determined using Eq.~(\ref{eq3}). 
In this calculation, the DT reconstruction efficiency, $\epsilon_{\rm{DT}}$, cannot be directly extracted from the DT MC samples because their $\cos\theta(\bar{n})$ distribution is inherited from \jnp~ and differs from that of \nppi. 
For large values of  $|\cos\theta(\bar{n})|$, the increased $\bar{n}$  penetration distance within the beam pipe enhances the likelihood of  $\bar{n}p$ scattering. 
To account for this, we derive $\epsilon_{\rm{DT}}(\cos\theta(\bar{n}))$ from DT MC and reweight it to match the  actual cos$\theta(\bar{n})$ distribution of the \nppi~ process.
After reweighting, the $\epsilon_{\rm{DT}}$ values decrease to 86\%--96\% of their uncorrected values, reflecting the greater difficulty in reconstructing tracks when $\bar{n}$ has a large absolute value of  $\cos\theta$.
The final values of $\epsilon_{\rm{DT}}$, along with the corresponding cross sections,  are presented in Table~\ref{table3}. 

\begin{table*}[!htbp]  
\caption{Summary of the measured $N_{\rm{DT}}$, $\epsilon_{\rm{DT}}$ (in \%), $\sigma_{i}$ and $\langle P \rangle$ for \nppi~ in full and different   $P(\bar{n})$ intervals. For $\sigma_{i}$, the first uncertainties are statistical, and the second systematic.}
\setlength{\tabcolsep}{2.0pt}
\centering 
\renewcommand{\arraystretch}{1.15}
\resizebox{\textwidth}{!}{
\begin{tabular}{c|c|c|c | c|c|c| c|c|c}
\hline \hline
\multirow{2}{*}{$P(\bar{n}) ~ / ~\langle P \rangle$ (MeV/$c$)}    &\multicolumn{3}{c|}{  $\bar{n}p \to 2\pi^{+}\pi^{-}$  }  &\multicolumn{3}{c|}{  $\bar{n}p \to 2\pi^{+}\pi^{-}\pi^{0}$   }   &\multicolumn{3}{c}{  $\bar{n}p \to 2\pi^{+}\pi^{-}2\pi^{0}$   }    \\   \cline{2-10}  
 &$N_{\rm{DT}}$ &$\epsilon_{\rm{DT}}$  & $\sigma_{0}$ (mb)  &$N_{\rm{DT}}$ &$\epsilon_{\rm{DT}}$ & $\sigma_{1}$ (mb)&$N_{\rm{DT}}$ &$\epsilon_{\rm{DT}}$ & $\sigma_{2}$ (mb) \\ \hline
  [200, 1174] /  765  &59.0 $\pm$ 7.7 &12.7 & 2.7 $\pm$ 0.3 $\pm$ 0.3        &182.0 $\pm$ 13.5 &9.8   &10.8 $\pm$ 0.8  $\pm$ 0.9          &97.7 $\pm$ 10.0 &4.6   &12.6 $\pm$  1.3  $\pm$ 1.0         \\ \hline
  $[200, 400]$ / 320 &3.0 $\pm^{2.1}_{1.4}$ &3.9   &5.6 $\pm^{3.9}_{2.7}$ $\pm$ 0.5 &16.0 $\pm^{4.3}_{3.7}$ &7.2    &16.7 $\pm^{4.5}_{3.9}$  $\pm$ 1.3  &11.0 $\pm^{3.7}_{3.0}$  &4.2   &19.9 $\pm^{6.6}_{5.4}$  $\pm$ 1.6 \\
  $[400, 600]$ / 503 &12.0 $\pm^{3.8}_{3.2}$ &9.7  &5.1 $\pm^{1.6}_{1.4}$ $\pm$ 0.5 &30.9 $\pm^{6.0}_{5.4}$ &9.7    &13.4 $\pm^{2.6}_{2.3}$  $\pm$ 1.1  &14.5 $\pm^{4.3}_{3.6}$  &4.7   &13.1 $\pm^{3.9}_{3.3}$  $\pm$ 1.1 \\   
  $[600, 800]$ / 710 &11.0 $\pm^{3.7}_{3.0}$ &13.2 &2.3 $\pm^{0.8}_{0.6}$ $\pm$ 0.2 &42.2 $\pm^{6.9}_{6.3}$ &10.4   &11.2 $\pm^{1.8}_{1.7}$  $\pm$ 0.9  &22.6 $\pm^{5.2}_{4.6}$  &4.7   &13.5 $\pm^{3.1}_{2.8}$  $\pm$ 1.1  \\   
  $[800, 1000]$ / 907 &21.0 $\pm^{4.9}_{4.3}$ &14.2 &2.5 $\pm^{0.6}_{0.5}$ $\pm$ 0.3 &64.0 $\pm^{8.3}_{7.7}$ &10.0   &10.8 $\pm^{1.4}_{1.3}$  $\pm$ 0.9  &29.1 $\pm^{5.8}_{5.1}$  &4.5   &11.0 $\pm^{2.2}_{1.9}$  $\pm$ 0.9  \\   
  $[1000, 1174]$ / 1063 &10.4 $\pm^{4.0}_{3.2}$ &15.1 &1.7 $\pm^{0.7}_{0.5}$ $\pm$ 0.2 &28.0 $\pm^{5.6}_{5.0}$ &10.1   &7.0  $\pm^{1.4}_{1.3}$  $\pm$ 0.6  &19.0 $\pm^{4.7}_{4.0}$  &4.6   &10.6 $\pm^{2.6}_{2.3}$  $\pm$ 0.9  \\ \hline   \hline
\end{tabular}\label{table3}
}
\end{table*}


The systematic uncertainty of the cross-section measurement is determined over the full momentum range and applied to the sub-momentum intervals of $\bar{n}$.
The systematic uncertainty related to the number of ST events is evaluated in the $RM(p\pi^{-})$ fit by varying the signal shape using the signal MC shape convolved with a Gaussian function with free parameters. The background shape is varied using a second-order Chebyshev polynomial function, and the fit range is changed by $\pm$0.01~GeV/$c^{2}$, resulting in a systematic uncertainty of 2.1\%. 
The systematic uncertainty due to the tracking or PID is 1\% per charged track~\cite{BES_scatter1}. For a single $\pi^{0}$ reconstruction, the systematic uncertainty is estimated to be 1\%~\cite{sys-pi0}.
Since the $\chi^{2}_{\text{kinematic}}$ requirement is loose ($<200$) in ST  but much more stringent  ($<5$)  in DT, we adjust the $\chi^{2}_{\text{kinematic}}$  requirement in  DT by $\pm$10\%.
The maximum deviation from the nominal  $\sigma_{i}$ values is assigned as the systematic uncertainty.
The systematic uncertainty from the angular distribution of $\bar{n}$ is 2.2\%, determined by the difference in $l$ values from the PWA and phase space  \jnp~ angular distributions. 
To address the uncertainty arising from the reweighting of MC efficiency, we adjust the counts per bin during reweighting  by adding or subtracting 20\% to observe the change in efficiency. The larger change is taken as the systematic uncertainty.
By varying the $2\pi^+\pi^-$ vertex requirement by $\pm10\%$, the  most significant $\sigma_i$ deviation is taken as the uncertainty.
For the $\bar{n}p$ scattering position, we adjust the requirements on $|\sqrt{(x_{1}^{2}+y_{1}^{2})} - \sqrt{(x_{2}^{2}+y_{2}^{2})}|$ and $|z_{1}-z_{2}|$  by  $\pm10\%$, and take the larger difference as the systematic uncertainty.
By modifying the $P(p_\text{oil})$  requirement by $\pm10\%$,   the larger difference is taken  as the systematic uncertainty.
Similar to the uncertainty in the $RM(p\pi^{-})$ fit, the uncertainty in the $\Delta E$ fit also includes three parts:  the signal shape using a Gaussian function with free parameters,  which also accounts for potential bias from the fixed parameters in $P(\bar{n})$ intervals, the background shape using a second-order Chebyshev polynomial function, and the fit range varying by $\pm$0.01~GeV.
To assess the uncertainty arising from the position of the IP, the coordinate origin is adjusted from $(0, 0, 0)$ to $(0.2, 0.2, 0.2)$~cm. The difference in the $l$ value is less than 0.1\% and is negligible. 
We check the cos$\theta$ and momentum distributions of the final particles $2\pi^{+}\pi^{-}i\pi^{0}$ from data and signal MC sample, and correct the MC efficiency based on data. The differences observed before and after this efficiency correction are taken as the systematic uncertainties.
A summary of the  systematic uncertainties is presented in Table~\ref{table2}. The total systematic uncertainty is obtained by adding all the individual components in quadrature. The systematic uncertainties on the cross-section measurements  are  shown in Table~\ref{table3}.

 \begin{table}[!htb] 
\caption{The systematic uncertainties in the cross section measurements  (in \%). The term ``neg.'' indicates negligible. The symbol ``$-$'' indicates  not applicable.}
\centering
\begin{tabular}{c|ccc}
\hline \hline
    Source                                      & $i=0$   & $i=1$   & $i=2$   \\ \hline 
    Number of ST events                         & 2.1     & 2.1     & 2.1     \\ 
    $2\pi^{+}\pi^{-}$ tracking                  & 3.0     & 3.0     & 3.0     \\ 
    $2\pi^{+}\pi^{-}$ PID                       & 3.0     & 3.0     & 3.0     \\
    $\pi^{0}$  reconstruction                   &  $-$    & 1.0     & 2.0     \\  
    $p\pi^{-}$  kinematic fit  $\chi^{2}$ cut   & 3.6     & 1.5     & 0.6     \\
    Angular distribution of $\bar{n}$           & 2.2     & 2.2     & 2.2     \\  
    Reweight MC efficiency                      & 0.2     & 0.2     & 0.2     \\ 
    $2\pi^{+}\pi^{-}$ vertex fit $\chi^{2}$ cut & 4.5     & 2.7     & 3.4     \\ 
    $\bar{n}p$ scattering position              & 1.3     & 0.5     & 1.2     \\  
    $P(p_\text{oil})$ requirement               & 1.0     & 3.1     & 3.7     \\ 
     $\Delta E$ fit strategy                    & 0.2     & 0.9     & 1.4     \\ 
    IP  position                                & neg.    & neg.    & neg.    \\ 
    Intermediate states                         & 2.6     & 2.6     & 1.8     \\ \hline  
              Total                             & 8.4     & 7.5     & 8.0     \\ \hline \hline
\end{tabular}\label{table2}
\end{table}

\begin{figure}[htbp]
\begin{center}
\begin{minipage}[t]{1.0\linewidth}
\includegraphics[width=0.90\textwidth]{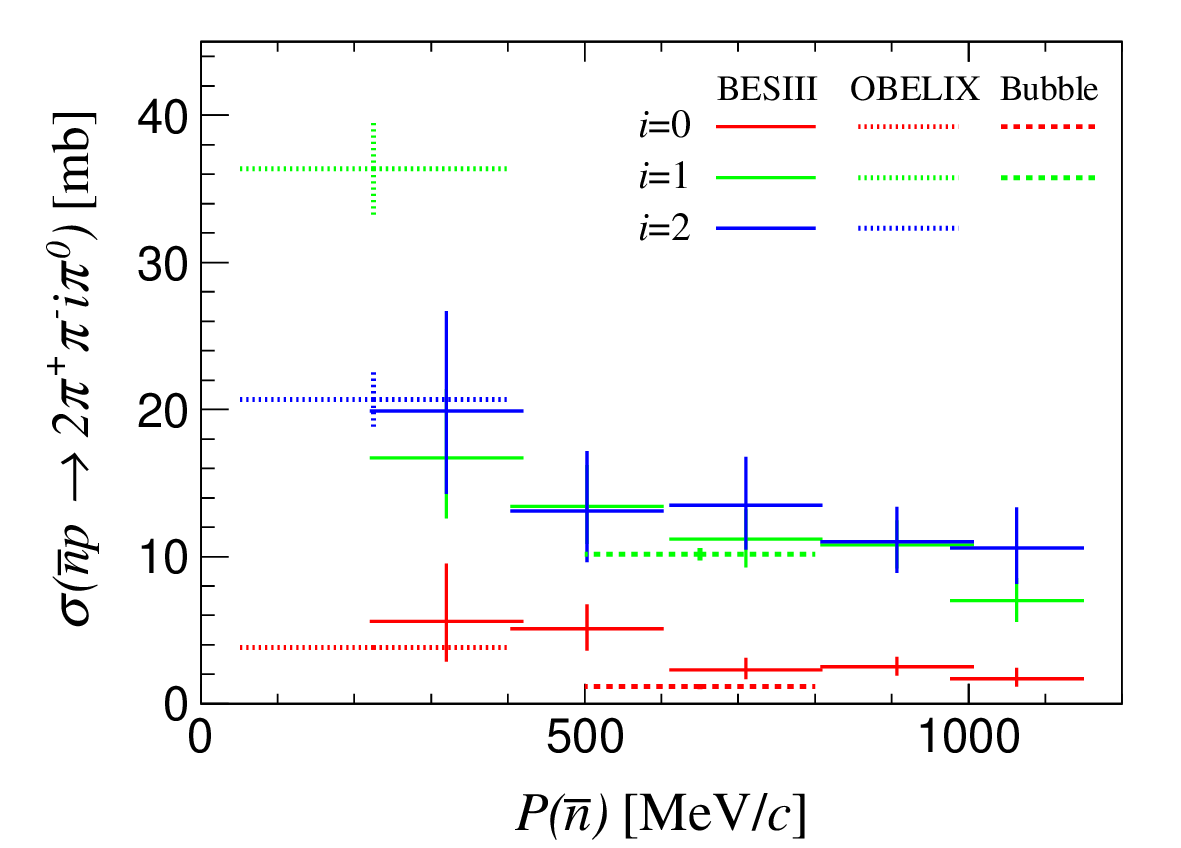}
\end{minipage}
\caption{Cross sections for \nppi~ (red for $i=0$, green for $i=1$, and blue for $i=2$). BESIII results are shown as solid lines with crosses in $P(\bar{n})$ intervals with the center of the bin indicated $\langle P \rangle$.  Vertical error bars show combined statistical and systematic uncertainties, and horizontal error bars indicate the $P(\bar{n})$ bin widths. Results from OBELIX and the bubble chamber experiment are shown as dotted and dashed lines, respectively, for comparison.} \label{line}
\end{center}
\end{figure}


    In summary, based on a sample of  $(10.087 \pm 0.044) \times 10^{9}$ $J/\psi$  events collected with the BESIII detector at the BEPCII storage ring, we report the first measurements of the reactions $\bar{n} p \to 2\pi^{+}\pi^{-}$, $\bar{n} p \to 2\pi^{+}\pi^{-}\pi^{0}$, and $\bar{n} p \to 2\pi^{+}\pi^{-}2\pi^{0}$, where the $\bar{n}$ is produced via the decay \jnp~ and the $p$ is  sourced from  hydrogen nuclei in the cooling oil of the beam pipe. 
    The cross sections for these reactions  are measured in five intervals of $P(\bar{n}):$ [200, 400], [400, 600], [600, 800], [800, 1000], and [1000, 1174]~MeV/$c$, within  the overall range of [200, 1174]~MeV/$c$, providing the first experimental data for the momentum range exceeding 800~MeV/$c$. 
    As shown in Fig.~\ref{line}, our interval cross section  results are compared with previous measurements from the OBELIX experiment\cite{allReview} for $i=0, 1, 2$ in the range [50, 400]~MeV/$c$ and a bubble chamber experiment~\cite{bubble} for $i=0, 1$ in the range [500, 800]~MeV/$c$. 
    Direct comparisons are limited by variations in the $\bar{n}$ momentum distributions across experiments.
    In the absence of direct theoretical predictions, we compare the relative channel strengths with the charge-conjugate $\bar{p}n \to 2\pi^{-}\pi^{+}i\pi^{0}$ annihilation at rest in liquid deuterium~\cite{pnD2}, where branching fractions of $(1.57\pm0.21)\%$, $(21.8\pm2.2)\%$, and $(6.3\pm1.1)\%$ are reported for $i=0, 1, 2$, respectively.
    Interestingly, while the $i=1$ channel dominates in $\bar{p}n$ annihilation at rest and in the lower-momentum OBELIX data, our measurements at higher momenta show that the  $\sigma_2$ is comparable to or exceeds $\sigma_1$ across most of the measured momentum range. This enhancement of multi-pion final states at higher energies provides new insights into the dynamics of nucleon-antinucleon annihilation.
    Moreover, intermediate $\rho$ and $\omega$ resonances are clearly observed in the invariant mass distributions of the $\pi\pi$ and $\pi\pi\pi$ subsystems, indicating the significant role of vector mesons in these annihilation processes. 
The reported results represent the first measurement of antineutron-nucleon inelastic scattering at an electron-positron collider, demonstrating the feasibility of using this  configuration to study such reactions. This research opens a path for future investigations at the  proposed super tau-charm factory~\cite{stcf1,stcf2}, where the abundant $J/\psi$ sample can provide an excellent source of antineutrons and long-lived (anti-)baryons for  high-precision studies in nuclear and particle physics.


The BESIII Collaboration thanks the staff of BEPCII (https://cstr.cn/31109.02.BEPC) and the IHEP computing center for their strong support. This work is supported in part by National Key R\&D Program of China under Contracts Nos. 2023YFA1606000, 2023YFA1606704; National Natural Science Foundation of China (NSFC) under Contracts Nos. 11635010, 11935015, 11935016, 11935018, 12025502, 12035009, 12035013, 12061131003, 12192260, 12192261, 12192262, 12192263, 12192264, 12192265, 12221005, 12225509, 12235017, 12361141819; the Chinese Academy of Sciences (CAS) Large-Scale Scientific Facility Program; the Strategic Priority Research Program of Chinese Academy of Sciences under Contract No. XDA0480600; CAS under Contract No. YSBR-101; 100 Talents Program of CAS; Program of Science and Technology Development Plan of Jilin Province of China under Contract No. 20230101021JC; The Institute of Nuclear and Particle Physics (INPAC) and Shanghai Key Laboratory for Particle Physics and Cosmology; ERC under Contract No. 758462; German Research Foundation DFG under Contract No. FOR5327; Istituto Nazionale di Fisica Nucleare, Italy; Knut and Alice Wallenberg Foundation under Contracts Nos. 2021.0174, 2021.0299; Ministry of Development of Turkey under Contract No. DPT2006K-120470; National Research Foundation of Korea under Contract No. NRF-2022R1A2C1092335; National Science and Technology fund of Mongolia; Polish National Science Centre under Contract No. 2024/53/B/ST2/00975; STFC (United Kingdom); Swedish Research Council under Contract No. 2019.04595; U. S. Department of Energy under Contract No. DE-FG02-05ER41374




\end{document}